\documentclass[slac_one]{revtex4}
\usepackage{graphicx}
\usepackage{fancyhdr}
\def\ifmath#1{\relax\ifmmode #1\else $#1$\fi}%
\def\TeV{\ifmmode {\,\mathrm{ Te\kern -0.1em V}}\else
                   \textrm{Te\kern -0.1em V}\fi}%
\def\GeV{\ifmmode {\,\mathrm{ Ge\kern -0.1em V}}\else
                   \textrm{Ge\kern -0.1em V}\fi}%
\def\MeV{\ifmmode {\,\mathrm{ Me\kern -0.1em V}}\else
                   \textrm{Me\kern -0.1em V}\fi}%
\def\keV{\ifmmode {\,\mathrm{ ke\kern -0.1em V}}\else
                   \textrm{ke\kern -0.1em V}\fi}%
\def\eV{\ifmmode  {\,\mathrm{ e\kern -0.1em V}}\else
                   \textrm{e\kern -0.1em V}\fi}%

\newcommand{\fb}{\,{\rm fb}}

\newcommand{\fbi}{\, \fb^{-1}}

\newcommand{\ee}    {\mathrm{e}^+\mathrm{e}^-}
\newcommand{\mumu}  {\mu^+\mu^-}

\newcommand{\qq}    {{\rm q\bar{\rm q}}}

\newcommand{\MZ}      {m_{\mathrm{Z}}}

\newcommand{\rts}   {\sqrt{s}}
\newcommand{\spr}  {\sqrt{s'}}

\begin{document}
\title{{\small{2005 International Linear Collider Workshop - Stanford,
U.S.A.}}\\ %% Please keep this conference title here
\vspace{12pt}
Measuring the Beam Energy with Radiative Return Events}
\author{A. Hinze, K. M\"onig}
\affiliation{DESY, Zeuthen, Germany}
\begin{abstract}
This paper studies the possibility to measure the centre of mass energy using
$\ee \rightarrow {\rm Z} \gamma \rightarrow \mumu \gamma$ events at the ILC.
With ${\cal L} = 100 \fbi$ at $\rts = 350 \GeV$ a relative error of around
$10^{-4}$ is possible. The potentially largest systematic uncertainty comes
from the knowledge of the aspect ratio of the detector.
\end{abstract}


\maketitle

\thispagestyle{fancy}
\section{INTRODUCTION}
The beam energy at the ILC is needed to a precision of around $10^{-4}$ for
accurate mass determinations of the top quark, the Higgs boson and in
supersymmetry \cite{tdr_phys}. This measurement will mainly be done with a
magnetic spectrometer \cite{spectrometer}. However the absolute calibration of
such a spectrometer is difficult and in addition the luminosity weighted
centre of mass energy is not necessarily identical to twice the beam energy.
It is thus very useful to have a method to measure the luminosity weighted
centre of mass energy directly from annihilation data. Fortunately this is
possible using radiative return events to the Z using the fact that the Z mass
is known to very high precision. The method was already pioneered at LEP2,
where it was however limited by the available statistics 
\cite{rrlepd,rrlepl,rrlepo}.

\section{ENERGY BIAS FROM THE KINK-INSTABILITY}
Wakefields in the main accelerator introduce a correlation between the z
position of an electron in the bunch and its energy. Due to the disruption of
the bunch in the interaction not all parts of the bunch contribute with the
same weight to the luminosity. The combination of both effects introduces a
bias in the luminosity weighted centre of mass energy. A detailed study can be
found in \cite{kink}.

\begin{figure}[htbp]
  \centering
  \includegraphics[width=0.45\linewidth]{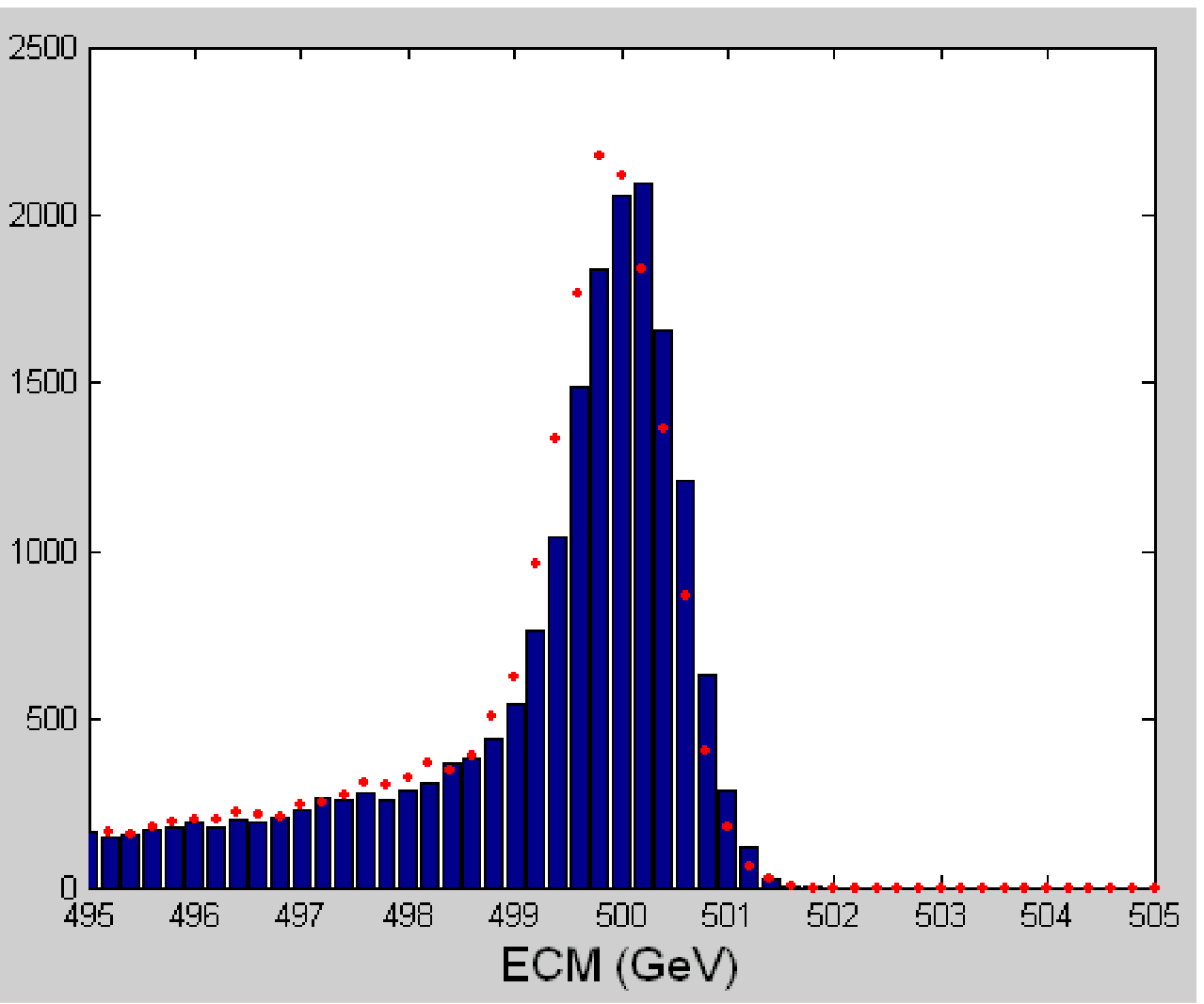}
  \includegraphics[width=0.45\linewidth]{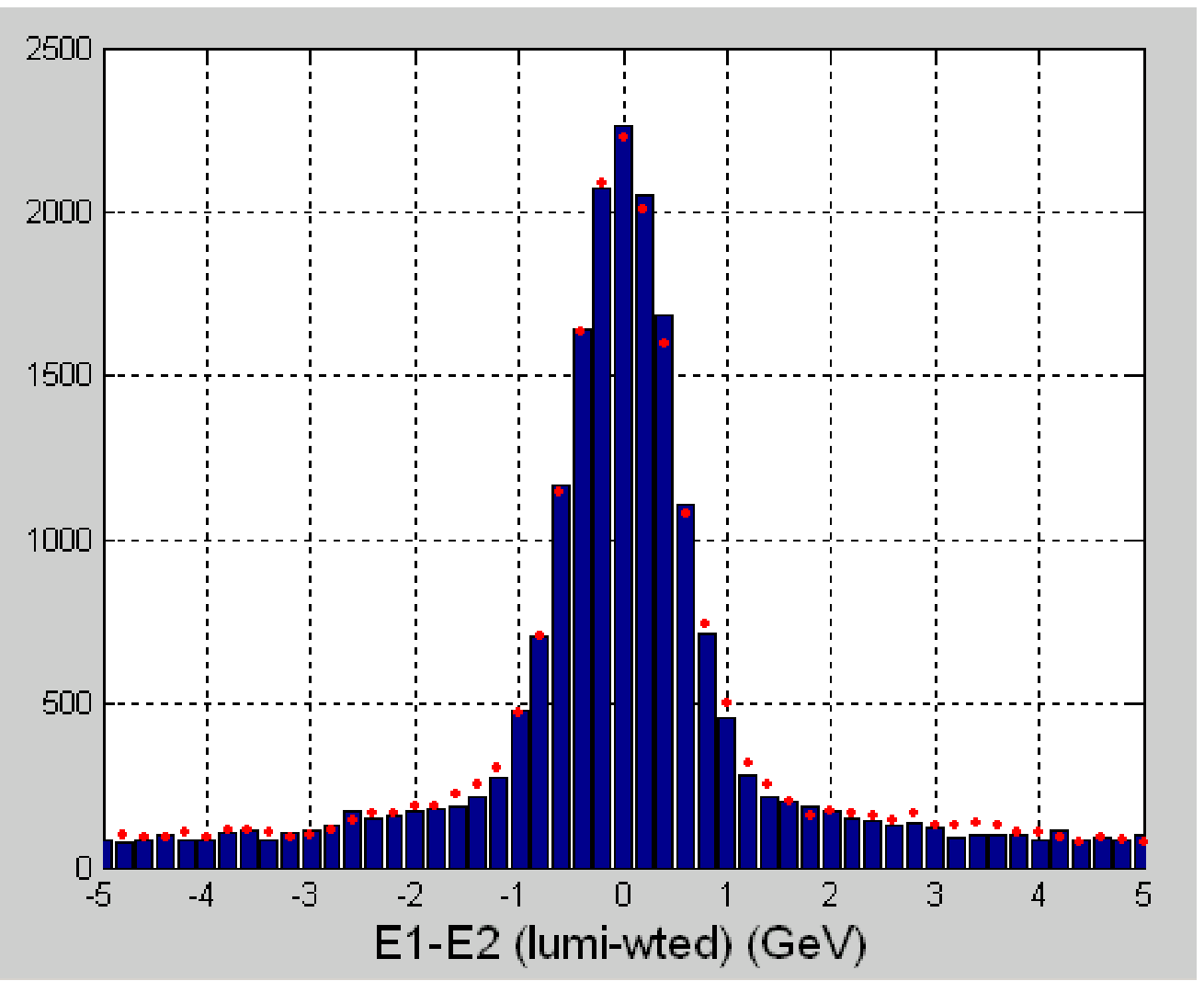}
  \caption{Centre of mass energy (left) and energy difference (right) of
  colliding electron-positron pairs in the TESLA design. The histogram shows
  the real simulation while for the dots the particle energies are ordered
  randomly in the bunch.}
  \label{fig:esumdif}
\end{figure}

For the TESLA design the effect is on average 150\,ppm with a spread of
30\,ppm and a maximum of 350\,ppm which is on the edge of being relevant.
Figure \ref{fig:esumdif} shows the centre of mass energy and the energy
difference of colliding particles for TESLA. The histogram shows the real
simulated distribution while the points show the artificial case where the
energies have been ordered randomly. The bias can be seen from the shift of
the mean of the two distributions in the centre of mass energy. The two
distributions agree well in the energy difference. This means that the bias
cannot be measured using the Bhabha acolinearity, which is proposed to measure
the beam energy spectrum due to beamstrahlung \cite{bhac}.
If one does not want to rely completely on beam simulations methods using
annihilation data are thus the only way to control such effects.

\section{THE RADIATIVE RETURN METHOD}
\begin{figure}[htbp]
  \centering
  \includegraphics[width=0.45\linewidth,bb=10 3 557 476]{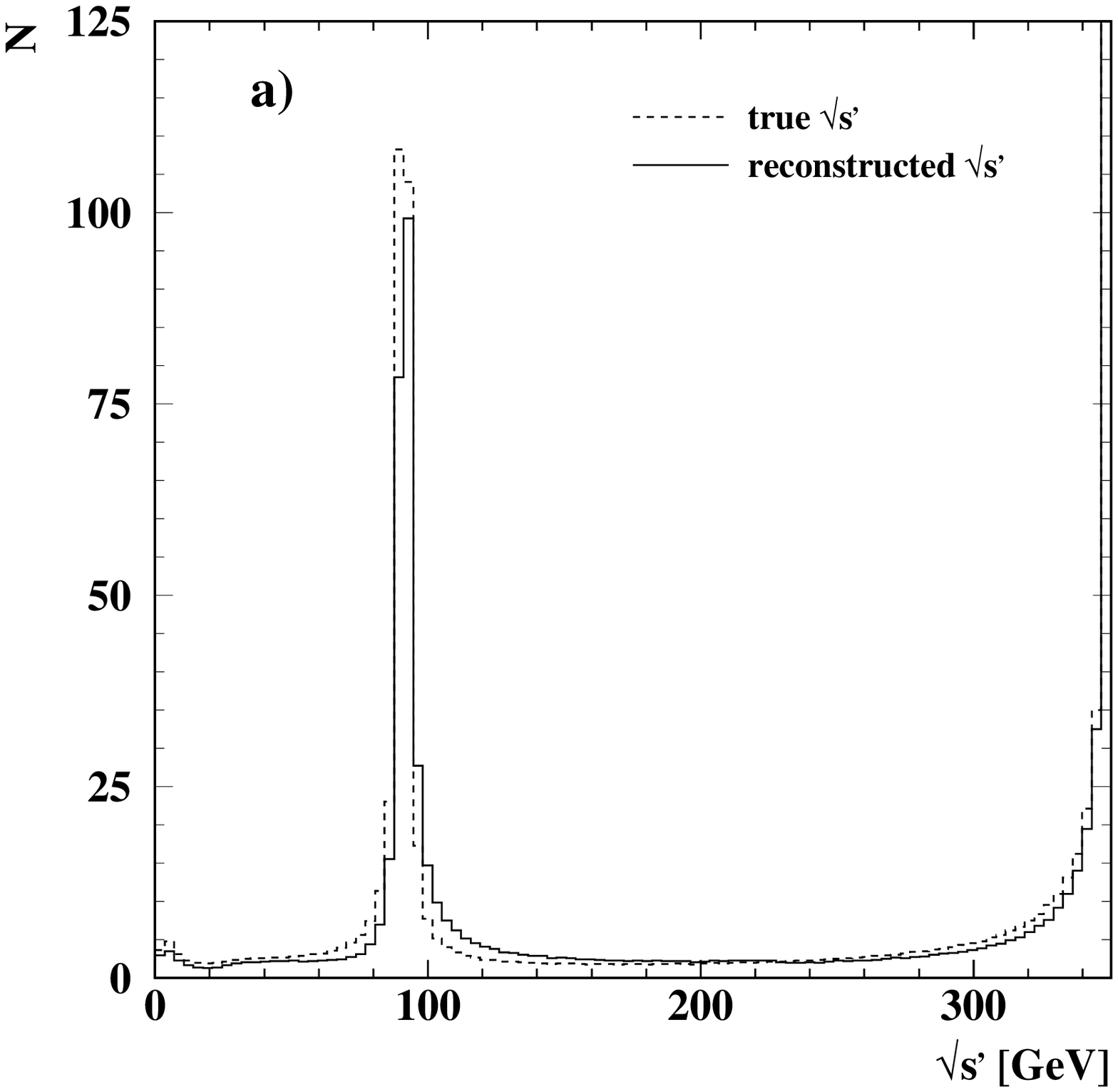}
  \includegraphics[width=0.45\linewidth,bb=10 3 557 476]{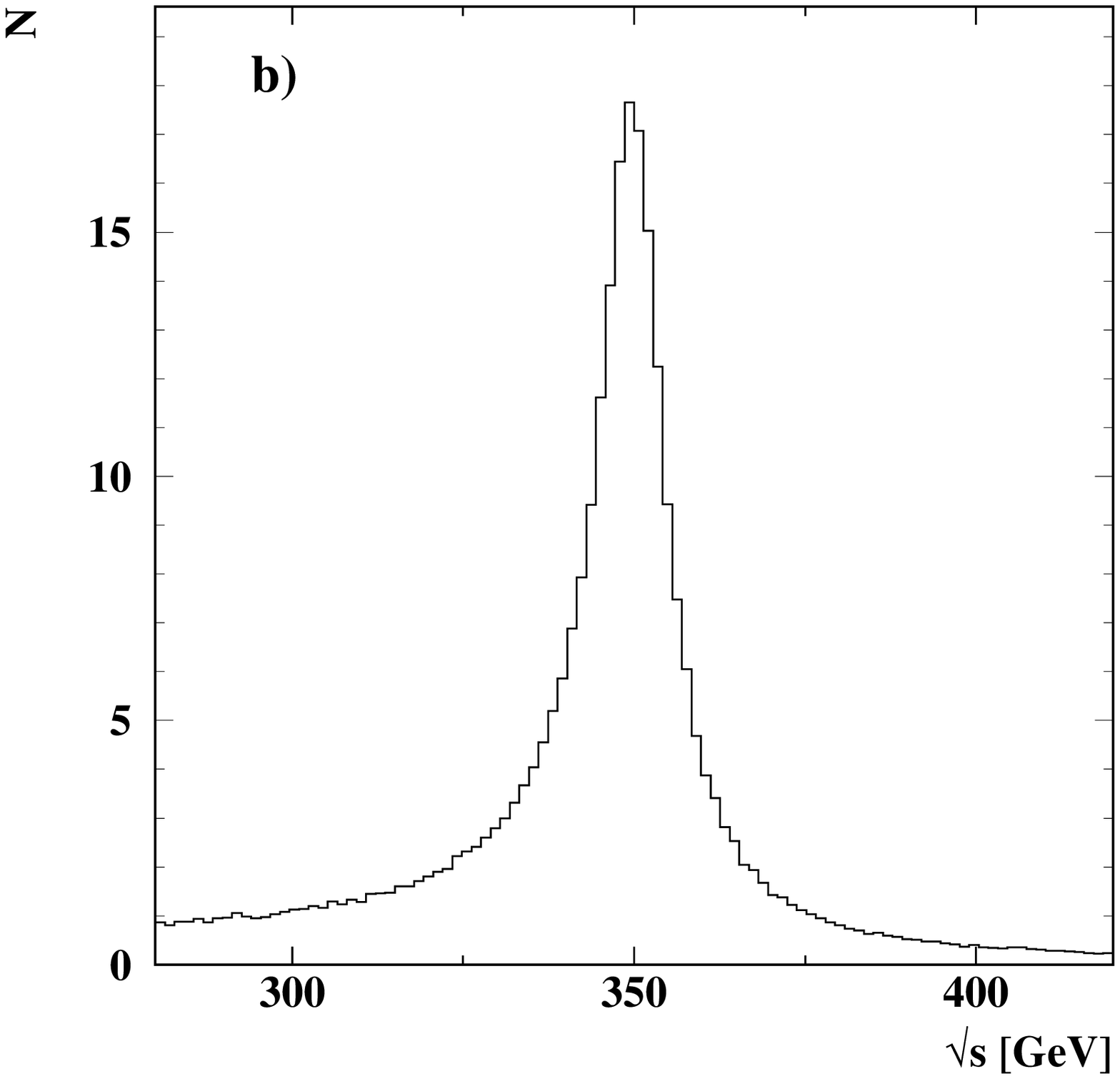}
  \caption{True and reconstructed $\spr$ (a) and reconstructed $\sqrt{s}$ for 
    $\ee \rightarrow {\rm Z} \gamma \rightarrow \mumu \gamma$ at $\sqrt{s}=350
    \GeV$}
  \label{fig:spr}
\end{figure}

The process $\ee \rightarrow {\rm Z} \gamma \rightarrow \mumu \gamma$ is well
suited for the reconstruction of of the centre of mass energy since the Z mass
is well known from LEP \cite{mzlep} and thus the $\gamma$ energy depends only
on the centre of mass energy, $\sqrt{s}$. If one assumes that exactly one
photon is radiated and that the energy of the two beams is the same, the mass
of the $\mumu$ system, $\spr$, can be reconstructed only from the angles of
the particles neglecting all energy measurements:
\begin{equation}
  \frac{\spr}{\sqrt{s}} = \sqrt\frac
               {\sin \theta_1 + \sin \theta_2 + \sin(\theta_1 + \theta_2)}
               {\sin \theta_1 + \sin \theta_2 - \sin(\theta_1 + \theta_2)}
\label{eq:spr}
\end{equation}
where $\theta_{1,2}$ are the angles between the two muons and the photon. In
most cases the photon is lost in the beampipe. In this case the photon
direction can be replaced by the z-axis signed by the negative $\mumu$
momentum vector.  In addition it is assumed that the fermion mass can be
neglected. Setting $\spr = \MZ$ one gets
\begin{equation}
\sqrt{s} = \MZ \sqrt\frac
               {\sin \theta_1 + \sin \theta_2 - \sin(\theta_1 + \theta_2)}
               {\sin \theta_1 + \sin \theta_2 + \sin(\theta_1 + \theta_2)}.
\label{eq:srec}
\end{equation}
Equation \ref{eq:srec} thus allows in principle to reconstruct the beam energy
without measuring energies with the detector. Only angles, which can be
measured with better precision and less systematic uncertainties are used.
In reality it is possible that more than one photon is radiated or
that one or both beams have lost energy due to beamstrahlung. These effects
can easily be accounted for in the fit, however they have to be known
accurately.
Figure \ref{fig:spr} a) shows the true $\mumu$ invariant mass and the one
reconstructed according to equation \ref{eq:spr} for $\sqrt{s} =350 \GeV$. 
Multiple radiation and
beamstrahlung is responsible for the shift of the Z-peak towards higher
energy. The reconstructed centre of mass energy is shown in figure
\ref{fig:spr} b). 

The cross section 
$\sigma(\ee \rightarrow {\rm Z} \gamma \rightarrow \mumu \gamma)$
is about 0.5\,pb for $\sqrt{s} =350 \GeV$ and scales approximately like
1/s. The detector accepts charged particles above $\theta = 7^\circ$
\cite{tdr_det} which
results in an efficiency of about 90\%.
For the simulation an ideal beam with a Gaussian energy spread of 0.2\% and
the CIRCE parameterisation of the beamstrahlung \cite{circe} has been used
\cite{arnd}.
\section{Background}
Potential backgrounds are given by all events that have exactly two muons in
the detector. These are
\begin{itemize}
\item two photon events and $\ee \rightarrow Z \ee$ events where the electrons
  are lost below $\theta = 7^\circ$;
\item $\ee \rightarrow {\rm ZZ}$ events where one Z decays into muons and the
  other into neutrinos;
\item $\ee \rightarrow {\rm W}^+ {\rm W}^-$ events where both W-bosons decay
  into a muon and a neutrino.
\end{itemize}
If there is no resonant Z-boson in the event the background can be rejected
efficiently by a cut around the reconstructed $\mumu$ mass, where the cut has
to be sufficiently loose not to reintroduce a dependence on the energy
calibration. For the analysis presented here a cut 
$\MZ - 5 \GeV < m(\mumu) < \MZ + 5 \GeV$ has been applied. Events with
neutrinos (WW, ZZ) can in principle be rejected further by a cut on the
tranverse momentum ballance, however this has been found to be not necessary.
The final sample contains a background of around 10\% from two-photon events,
25\% from Zee events and about 1\% from WW and ZZ. The Zee background is
rather large, but this does nor pose any problem. The typical topology for
these events is one electron of very high energy while the momentum of the
other one is very low. These events thus have a sensitivity to the beam energy
very similar to the signal events.

\section{Fit Method}
To fit the beam energy from the reconstructed centre of mass energy a Monte
Carlo linearising around a default value \cite{bhac} has been used. In this
method it is assumed that the differential cross section at a given
reconstructed centre of mass energy is a linear function of the true centre of
mass energy in a range around the nominal value larger than the expected
error:
\begin{eqnarray*}
\sigma(\rts,\rts_{\rm rec}) & = & \sigma(\rts_0,\rts_{\rm rec})
+ A(\rts_0,\rts_{\rm rec}) \left( \rts - \rts_0 \right) \\
A(\rts_0,\rts_{\rm rec}) & = & 
\frac{\sigma(\rts_1,\rts_{\rm rec})-\sigma(\rts_0,\rts_{\rm rec})}
{\rts_1 - \rts_0}.
\end{eqnarray*}
$\sigma(\rts_i,\rts_{\rm rec}),\, i=0,1$ is calculated with the simulation
including all effects like background, detector resolution, beamstrahlung etc.
Apart from the linearity assumption the fit is bias free per construction and
this assumption can be tested with the simulation to be valid.  

In the fit the data a binned in $\rts_{\rm rec}$ and a $\chi^2$ is built as a
function of $\rts$, summing over the bins in $\rts_{\rm rec}$. Not to be
dependent on the luminosity measurement the total normalisation was treated as
a second free parameter in the fit.

\section{Results}

Monte Carlo data corresponding to an integrated luminosity of
${\cal L} = 100 \fbi$ at $\rts = 350 \GeV$ have been fitted with the method
described above. Including background, beamstrahlung and energy spread an
error of $\Delta \sqrt{s} = 47 \MeV$ or 
$\frac{\Delta \sqrt{s}}{\sqrt{s}} = 1.3\cdot 10^{-4}$ has been achieved.
If beamstrahlung and energy spread are omitted the error is about 10\%
smaller. The influence of the background is negligible.

It has been shown that this error can be improved by a factor two to four if
the muon momenta are included in the fit \cite{tim}. For this improvement a
momentum resolution with a constant term of around $2 \cdot 10^{-5}/\GeV$ and a
multiple scattering term of around $10^{-3}$ is needed. Furthermore it is
assumed that the systematic uncertainty on the momentum resolution can be
described by a single scale factor which is included as a free parameter in
the fit.

\begin{figure}[htbp]
  \centering
  \includegraphics[width=0.5\linewidth,bb=10 3 557 476]{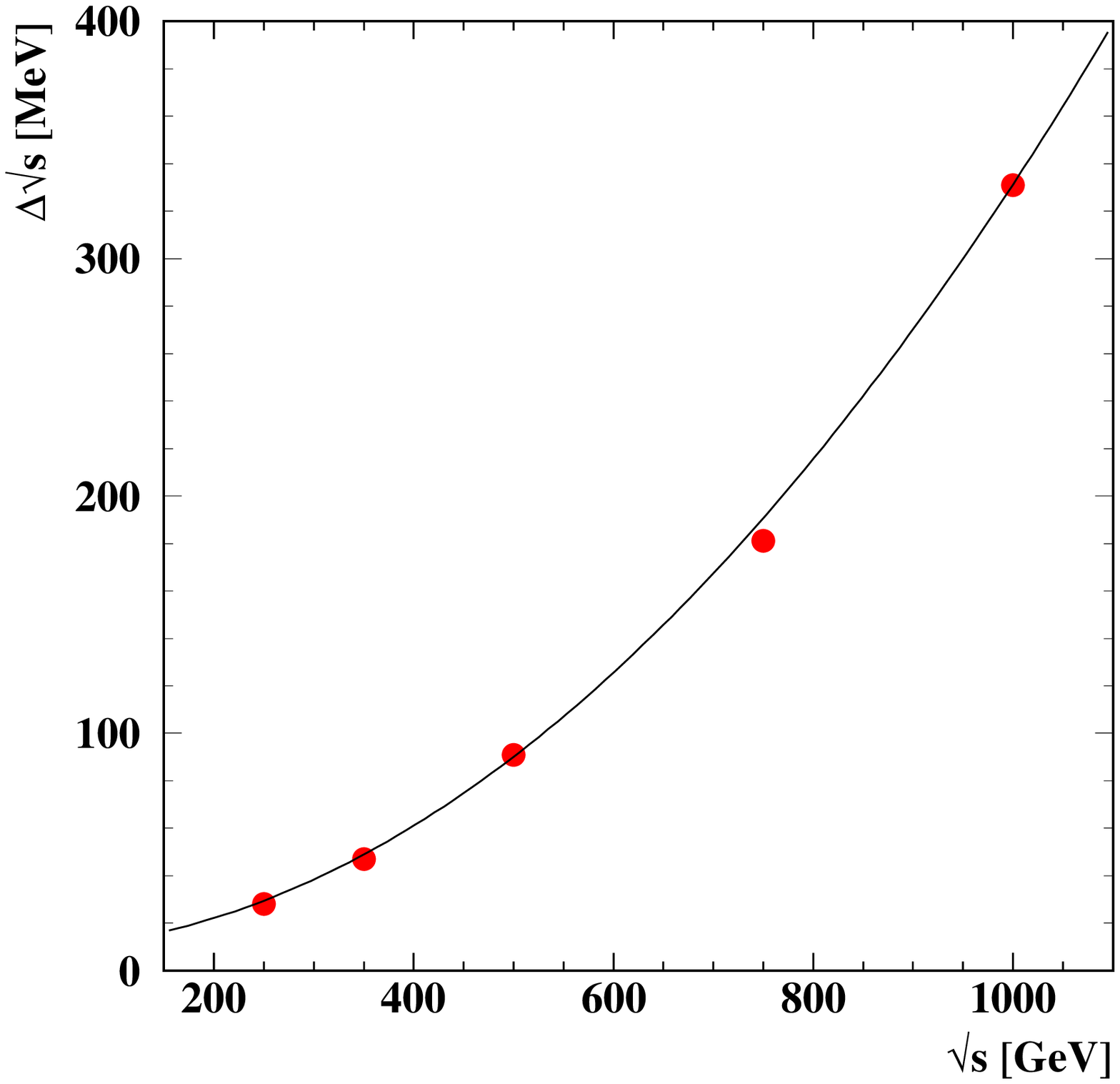}
  \caption{Energy dependence of $\Delta \rts$ for ${\cal L} = 100 \fbi$.}
  \label{fig:ecm}
\end{figure}

As shown in figure \ref{fig:ecm} the error depends strongly on the centre of
mass energy. For constant luminosity the error can be parametrised as 
\[
\Delta \sqrt{s} = \left( 8.8+0.0026 \sqrt{s}/\GeV +0.0032 s/\GeV^2\right) \MeV.
\]
It should, however, be noted that the relative error is almost constant if the
luminosity increases proportional to $s$.

Several sources of systematic uncertainty have been studied. The background
has no effect if an uncertainly of less than 20-30\% on the amount of
background is assumed. If instead of a Gaussian energy spread a rectangular
shape is assumed the reconstructed centre of mass energy changes by
$10\MeV$. There is no change if the width is changed from 0.2\% to 0.1\%.

If the parameters describing the beamstrahlung in Circe are varied by values as
suggested in \cite{bhac} a shift of the beam energy up to $40 \MeV$ has been
found. This shift is, however, strongly anticorrelated with the shift of the
mean beam energy due to the parameter change, so that the uncertainty on the
average beam energy is very small.

The by far largest error may come from an uncertainty in the polar angle
measurement of the detector. At LEP is was assumed that the ratio of the
detector radius and detector length is 
$\Delta \left(\frac{R}{L}\right) = \Delta \tan \theta
= 5 \cdot 10^{-4}$. If the same uncertainty hold for the ILC detector the
uncertainty on the reconstructed centre of mass energy would be 
$\Delta \sqrt{s} = 160 \MeV$. The aspect ratio of the detector thus needs to
be known an order of magnitude more accurate than at LEP to make the beam
energy measurement with radiative return events useful.

\subsection{Future Work}

It would be useful to increase the statistics of the radiative return
measurement.  Bhabha scattering ($\ee \rightarrow \ee$) is in principle clean,
however the signal is diluted by the t-channel contribution. However with a
cut on the production angle in the centre of mass system a useful measurement
should still be possible.  The resolution for $\tau^+ \tau^-$ events will be
somewhat diluted because of the kink of the charged particles in the $\tau$
decay. The main problem, however, is that due to the missing neutrinos the cut
on the $\tau^+ \tau^-$ invariant mass to reject two-photon background is not
very effective.

In principle there is a much larger statistics using $Z\rightarrow \qq$
events.
As already said, equation \ref{eq:srec} assumes, however, that the mass of the
final state particle is negligible. A $5\GeV$ jet mass results in a shift of
$2.5 \GeV$ in $\spr_{\rm rec}$. It is thus very improbable that fragmentation
can be understood well enough to make these events useful.

To get a final estimate of the radiative return method a global analysis will
be needed. Beamstrahlung and the kink instability are correlated between the
two beams. These correlations influence the Bhabha acolinearity to measure the
beamstrahlung and the reconstructed $\rts$ from the radiative return analysis
simultaneously. A common analysis using both methods is thus needed to see how
these effects modify the reconstructed centre of mass energy.

\section{Conclusions}
The centre of mass energy can be measured on the $10^{-4}$ level from
radiative return events using only the measured angles of the final state
muons.  This is, however, a high luminosity analysis. The statistics is not
sufficient to measure $\rts$ for example point by point in a mass scan. These
relative measurements still have to be done using spectrometers.

The potentially largest systematic uncertainty comes from the aspect ratio of
the detector. Great care has to be taken in the detector design to make sure
that this quantity is understood on the $10^{-4}$ level.

To draw final conclusions on this method a global analysis of the acolinearity
of Bhabha events for beamstrahlung and of the radiative return events for the
beam energy is needed to understand the effects from beam-beam correlations.

%\begin{acknowledgments}
%Thanks to everybody
%\end{acknowledgments}

\end{document}